\documentclass[twocolumn]{aa}

\usepackage{euclid}
\usepackage{graphicx}
\usepackage{natbib}
\usepackage{scalerel}
\usepackage{physics}
\usepackage{tikz}

\usepackage{txfonts}
\usepackage[pdfencoding=auto,psdextra]{hyperref}
\hypersetup{
    colorlinks=true,
    linkcolor=blue,
    filecolor=magenta,      
    urlcolor=blue,
    citecolor=blue
}
\makeatletter
\renewcommand*\aa@pageof{, page \thepage{} of \pageref*{LastPage}}
\makeatother

\usepackage[utf8]{inputenc}

\usepackage[switch, modulo]{lineno}

\usepackage{float} 

\newcommand{\Omm}{\Omega_\mathrm{m}}
\newcommand{\Map}{M_\mathrm{ap}}
\newcommand{\MapEst}{\hat{M}_\mathrm{ap}}

\newcommand{\MapMapMap}{\expval{\Map^3}}
\newcommand{\MapMap}{\expval{\Map^2}}

\newcommand{\astroang}[1]{\ang[angle-symbol-over-decimal]{#1}}
\newcommand{\ellvec}{\vec{\ell}}
\newcommand{\dirac}{\delta_\mathrm{D}}

\newcommand{\varthetavec}{\vec{\vartheta}}

\newcommand{\I}{\mathrm{i}}
\newcommand{\E}{\mathrm{e}}
\newcommand{\etavec}{\vec{\eta}}
\newcommand{\alphavec}{\vec{\alpha}}
\newcommand{\qvec}{\vec{q}}
\newcommand{\realspace}{\mathbb{R}}

\newcommand{\MapMapEst}{\MapEst^2}

\begin{document}

\title{What is the super-sample covariance? A fresh perspective for second-order shear statistics.}

\newcommand{\orcid}[1]{} 

\author{Laila Linke \inst{1,2}\thanks{Corresponding author, \email{laila.linke@uibk.ac.at}},
        Pierre A. Burger \inst{1},
        Sven Heydenreich \inst{1,3},
        Lucas Porth \inst{1},
        Peter Schneider \inst{1}
      }

\institute{
      \inst{1} Argelander-Institut f\"ur Astronomie, Auf dem H\"ugel 71, 53121 Bonn, Germany \\
      \inst{2} Universit\"at Innsbruck, Institut f\"ur Astro- und Teilchenphysik, Technikerstr. 25/8, 6020 Innsbruck, Austria\\
      \inst{3} Department of Astronomy and Astrophysics, University of California, Santa Cruz, 1156 High Street, Santa Cruz, CA 95064 USA
      }

\abstract{Cosmological analyses of second-order weak lensing statistics require precise and accurate covariance estimates. These covariances are impacted by two sometimes neglected terms: a negative contribution to the Gaussian covariance due to a finite survey area, and the super-sample covariance (SSC), which for the power spectrum contains the impact of Fourier modes larger than the survey window. We show here that these two effects are connected and can be seen as correction terms to the `large-field-approximation', the asymptotic case of an infinitely large survey area. We describe the two terms collectively as `Finite-Field-Terms'.

We derive the covariance of second-order shear statistics from first principles. For this, we use an estimator in real space without relying on an estimator for the power spectrum. The resulting covariance does not scale inversely with the survey area, as might naively be assumed. This scaling is only correct under the large-field approximation when the contribution of the finite-field terms tends to zero. Furthermore, all parts of the covariance, not only the SSC, depend on the power spectrum and trispectrum at all modes, including those larger than the survey. 
We also show that it is generally impossible to transform an estimate of the power spectrum covariance into the covariance of a real-space statistic. Such a transformation is only possible in the asymptotic case of the large-field approximation.

Additionally, we find that the total covariance of a real-space statistic can be calculated using correlation function estimates on spatial scales smaller than the survey window. Consequently, estimating covariances of real-space statistics, in principle, does not require information on spatial scales larger than the survey area. We demonstrate that this covariance estimation method is equivalent to the standard sample covariance method.}
%
%
%
%
\titlerunning{What is the SSC?}
\authorrunning{Linke et al.}
   
   \maketitle
%
%
%
%

\section{\label{sec: introduction}Introduction}
Second-order statistics of cosmic shear are essential tools for cosmological analyses (\citealp{Heymans2021}, \citealp{Hikage2019}, \citealp{Abbott2022}). Inference of cosmological parameters from these statistics requires a robust understanding of their covariances. While covariance models for second-order shear statistics have been derived and validated for over a decade \citep{Joachimi2008, Takada2013}, two effects have only recently garnered more attention; specifically, the super-sample covariance (SSC) and the impact of the survey window on the Gaussian covariance. Our goal here is to show that these two effects are related and caused by the same approximation on the survey window function.

Analytic models for the covariance of second-order statistics are usually expressed in terms of the power spectrum covariance. The power spectrum covariance can be divided into three terms: a Gaussian and an intra-survey non-Gaussian part, which depend only on the power spectrum and trispectrum at $\ell$-modes within a survey area, and the SSC, which also depends on $\ell$-modes outside the survey \citep{Takada2013}. While the first two terms scale linearly with the inverse survey area, the SSC shows a complicated dependence on the survey window function \citep{Lacasa2019, GouyouBeauchamps2022}. An analytic model for the SSC was derived in \citet{Barreira2017}. Additionally, N-body simulations with small box sizes and periodic boundary conditions cannot fully reproduce the SSC of the matter power spectrum, as they do not include the small $\ell$-modes (large spatial scales) on which the SSC depends \citep{DePutter2012, Takahashi2009}. Instead, one needs to use either large boxes, from which only a small region is taken to estimate the power spectrum \citep[e.g.][]{Bayer2023}, or `separate universe simulations', where multiple realisations with varying mean densities are simulated \citep{Li2014}. As shown in \citet{Barreira2018}, the SSC is usually larger than the intra-survey non-Gaussian term and therefore the dominating contribution to the non-diagonal elements of the covariance. Estimating it correctly is therefore vital for cosmic shear analyses.

However, while the SSC for the power spectrum can be interpreted as capturing the clustering information at $\ell$-modes outside a survey, the same interpretation is not necessarily accurate for statistics in real space. These real-space statistics, such as shear correlation functions \citep{Kaiser1992, Amon2022} or complete orthogonal sets of E-/B-integrals (COSEBIs; \citealp{Schneider2010, Asgari2020}) are preferred for cosmological analyses, as they can be directly estimated from survey data. Here, we will derive the covariance for the estimator of general second-order shear statistics in real space to find an interpretation of the SSC of these statistics. We will show the following four key findings.

First, for a `localised'\footnote{see Eq.~\eqref{eq: definition second order statistic} for a definition; examples are the aperture statistics \citep{Schneider1998} and COSEBIs \citep{Schneider2010}} second-order shear statistic $\Xi$, the full covariance $C_{\hat{\Xi}}$ can be obtained from correlation functions of the convergence field smoothed according to the chosen statistic. These correlation functions need to be known only on scales smaller than the survey area. Correlations on spatial scales larger than the survey area do not impact the covariance, including the Gaussian finite-field and the SSC term.
    
Second, the exact covariances of both $\Xi$ and the power spectrum $P$ do not scale inversely with the survey area. Instead, they also depend on the survey geometry via the SSC and a suppression term of the Gaussian covariance term. Only under the assumption of an infinitely broad survey window function do these terms vanish and we recover the naive scaling with survey area.

Third, all parts of $C_{\hat{\Xi}}$ depend on the power spectrum and trispectrum at $\ell$-modes within and outside of the survey area. Therefore, the SSC term alone does not capture the clustering information at $\ell$-modes outside a survey.

Finally, it is, in general, not possible to transform the covariance of the power spectrum to the covariance of a real-space statistic. Such a conversion requires the assumption of an infinitely broad survey window.

The present paper is structured as follows: In Sect.~\ref{sec: power spectrum}, we discuss the covariance of the power spectrum and show the origin of the finite-field terms. In Sect.~\ref{sec: xi real space}, we introduce an estimator for a real-space statistic $\Xi$ and show how its covariance is related to correlation functions of $\Xi$. We demonstrate in Sect.~\ref{sec: cov corr} that using correlation function estimates gives the same result (up to a prefactor) as the usual sample covariance approach. In Sect.~\ref{sec: xi fourier space}, we connect the covariance of $\Xi$ to the power spectrum and trispectrum and show that the finite-field terms for $\Xi$ are given by the difference between the exact covariance and an approximation of the covariance for an infinitely broad survey window function. We conclude in Sect.~\ref{sec: conclusion}. 

Throughout this paper, we are working in the flat-sky limit. Any figures and calculations are performed using the parameters and simulations described in Appendix~\ref{app: validation}. We note that we are not explicitly giving the dependence of the covariances on shape noise as its effect can be included by replacing the power spectrum $P$ by $P+\sigma^2_\epsilon/2n$ in all covariance expressions, where $n$ is the galaxy number density and $\sigma^2_\epsilon$ the two-component ellipticity dispersion.

\section{\label{sec: power spectrum}Power spectrum covariance}

Before we consider the covariance of a real space statistic, we first give an overview of the covariance of the power spectrum and the origin of the SSC based on \citet{Takada2013}. We are considering here the power spectrum of the weak-lensing convergence, which is the normalised surface mass density, which itself is related to the density contrast $\delta$. In a flat universe, and at angular position $\varthetavec$ and comoving distance $\chi$, the convergence is
\begin{equation}
    \kappa(\varthetavec) = \frac{3H_0^2\Omm}{2c^2}\int_0^\infty \dd{\chi}\; q(\chi)\,\chi\,\frac{\delta(\chi\varthetavec, \chi)}{a(\chi)}\;,
\end{equation}
where
\begin{equation}
   q(\chi) = \int_\chi^\infty \dd{\chi'}\, p(\chi')\, \frac{\chi'-\chi}{\chi'}\;, 
\end{equation}
with the Hubble constant $H_0$, the matter density parameter $\Omega_\mathrm{m}$, the cosmic scale factor $a(\chi)$ at $\chi$, normalised to unity today, and the probability distribution $p(\chi)\,\dd{\chi}$ of source galaxies with comoving distance.

The convergence power spectrum $P(\ell)$ is defined by
\begin{equation}
    (2\pi)^2\, \dirac(\ellvec+\ellvec')\, P(\ell) = \expval{\tilde{\kappa}(\ellvec)\, \tilde{\kappa}(\ellvec')}\;,
\end{equation}
where $\kappa$ is the convergence and the tilde denotes Fourier transform. We assume a survey of simple geometry (i.e. continuous and without small-scale masks) with a window function $W$, which is either zero or one, and survey area $A$, given by $A=\int\dd[2]{\vartheta}\, W(\varthetavec)$. Here, $P$ can be estimated with the estimator
\begin{align}
\label{eq: power spectrum estimator}
    \hat{P}(\ell)&=\frac{1}{A}\int_{A_R(\ell)}\frac{\dd[2]{\ell}'}{A_R(\ell)} \left[\prod_{i=1}^2 \int \frac{\dd[2]{q_i}}{2\pi} \tilde{W}(\qvec_i)\right]\, \\
    &\notag \quad \times\tilde{\kappa}(\ellvec'-\qvec_1)\, \tilde{\kappa}(-\ellvec'-\qvec_2)\;,
\end{align}
where $A_R(\ell)$ denotes the size of the $\ell$-bin.

The usual form used for the covariance of $\hat{P}$ in the literature \citep{Takada2013, Krause2017} is
\begin{align}
\label{eq: Covariance literature}
    &\notag C_{\hat{P}}^\mathrm{lit}(\ell_1, \ell_2)\\
    &= \frac{2}{A}\int_{A_R(\ell_1)}\frac{\dd[2]{\ell'_1}}{A_R(\ell_1)}\int_{A_R(\ell_2)}\frac{\dd[2]{\ell'_2}}{A_R(\ell_2)}\, P(\ell'_1)\, P(\ell'_2)\, \dirac(\ellvec'_1+\ellvec'_2) \\
    &\notag \quad + \frac{1}{A}\int_{A_R(\ell_1)}\frac{\dd[2]{\ell'_1}}{A_R(\ell_1)}\int_{A_R(\ell_2)}\frac{\dd[2]{\ell'_2}}{A_R(\ell_2)}\, T(\ellvec'_1, -\ellvec'_1, \ellvec'_2, -\ellvec'_2)\\
    &\notag \quad + C_{\hat{P}}^\mathrm{SSC}(\ell_1, \ell_2)\;, 
\end{align}
 where $T$ is the convergence trispectrum and $C_{\hat{P}}^\mathrm{SSC}$ is the SSC of the power spectrum. \citet{Takada2013} derive the SSC to be
\begin{align}
    C_{\hat{P}}^\mathrm{SSC}(\ell_1, \ell_2) &= \frac{1}{A^2}\int_{A_R(\ell_1)}\frac{\dd[2]{\ell'_1}}{A_R(\ell_1)}\int_{A_R(\ell_2)}\frac{\dd[2]{\ell'_2}}{A_R(\ell_2)} \\
    &\notag \quad \times \int\frac{\dd[2]{q}}{(2\pi)^2} T_\mathrm{SSC}(\ellvec_1, \ellvec_2, \qvec)\, \tilde{W}(\qvec) \, \tilde{W}(-\qvec)\;,
\end{align}
where $T_\mathrm{SSC}$ is part of the convergence trispectrum and given by Equation (32) in \citet{Takada2013}. However, Eq.~\eqref{eq: Covariance literature} is incomplete. To see this, we start from Eq.~\eqref{eq: power spectrum estimator}, so $C_{\hat{P}}$ is
\begin{align}
   &C_{\hat{P}}(\ell_1, \ell_2)=\expval{\hat{P}(\ell_1)\, \hat{P}(\ell_2)}-\expval{\hat{P}(\ell_1)}\expval{\hat{P}(\ell_2)}\\
&\notag  = \frac{1}{A^2}\int_{A_R(\ell_1)}\frac{\dd[2]{\ell'_1}}{A_R(\ell_1)}\int_{A_R(\ell_2)}\frac{\dd[2]{\ell'_2}}{A_R(\ell_2)}\, \left[\prod_{i=1}^4 \int \frac{\dd[2]{q_i}}{(2\pi)^2}\,\tilde{W}(\qvec_i)\right] \\
&\notag \quad \times \left[ \expval{\Tilde{\kappa}(\ellvec'_1+\qvec_1)\,\Tilde{\kappa}(-\ellvec'_1+\qvec_2)\,\Tilde{\kappa}(\ellvec'_2+\qvec_3)\,\Tilde{\kappa}(-\ellvec'_2+\qvec_4)}\right.\\
&\notag \quad - \left. \expval{\Tilde{\kappa}(\ellvec'_1+\qvec_1)\,\Tilde{\kappa}(-\ellvec'_1+\qvec_2)}\,\expval{\Tilde{\kappa}(\ellvec'_2+\qvec_3)\,\Tilde{\kappa}(-\ellvec'_2+\qvec_4)} \right]
\end{align}
The four-point function of $\Tilde{\kappa}$ can be decomposed into its connected and unconnected parts and can be written in terms of the power spectrum and trispectrum as
\begin{align}
\label{eq: split kappa four point}
     &\notag\expval{\Tilde{\kappa}(\ellvec_1)\,\Tilde{\kappa}(\ellvec_2)\,\Tilde{\kappa}(\ellvec_3)\,\Tilde{\kappa}(\ellvec_4)} \\
     &= \expval{\Tilde{\kappa}(\ellvec_1)\,\Tilde{\kappa}(\ellvec_2)\,\Tilde{\kappa}(\ellvec_3)\,\Tilde{\kappa}(\ellvec_4)}_\mathrm{c}\\
     &\quad\notag+\left[ \expval{\Tilde{\kappa}(\ellvec_1)\,\Tilde{\kappa}(\ellvec_2)}\,\expval{\Tilde{\kappa}(\ellvec_3)\,\Tilde{\kappa}(\ellvec_4)} +\textrm{2 Perm.}\right]\\
     &\notag = (2\pi)^2\,T(\ellvec_1, \ellvec_2, \ellvec_3, \ellvec_4)\,\dirac(\ellvec_1+\ellvec_2+\ellvec_3+\ellvec_4)\\
     &\quad\notag + \left[ (2\pi)^4 P(\ell_1)\,P(\ell_3)\,\dirac(\ellvec_1+\ellvec_2)\,\dirac(\ellvec_3+\ellvec_4) +\textrm{2 Perm.}\right]\;.
\end{align}
Therefore, after suitably renaming of the $q_i$, using $\qvec=\qvec_1+\qvec_2$, and after evaluating the Dirac-functions and simplifying,
\begin{align}  
\label{eq: Covariance power spectrum unapproximated}
     C_{\hat{P}}(\ell_1, \ell_2)&=\notag\frac{2}{A^2}\int_{A_R(\ell_1)}\frac{\dd[2]{\ell'_1}}{A_R(\ell_1)}\int_{A_R(\ell_2)}\frac{\dd[2]{\ell'_2}}{A_R(\ell_2)}\\
    & \quad \times \int \frac{\dd[2]{q_1}}{(2\pi)^2} \int \frac{\dd[2]{q_2}}{(2\pi)^2} P(|\ellvec'_1-\qvec_1|)\, P(|\ellvec'_2+\qvec_2|)\, \\
    &\notag\quad\times \tilde{W}(\qvec_1)\, \tilde{W}(\qvec_2) \,\tilde{W}(\ellvec'_1+\ellvec'_2-\qvec_1)\, \tilde{W}(-\ellvec'_1-\ellvec'_2-\qvec_2)\\
    &\notag+\frac{1}{A^2}\int_{A_R(\ell_1)}\frac{\dd[2]{\ell'_1}}{A_R(\ell_1)}\int_{A_R(\ell_2)}\frac{\dd[2]{\ell'_2}}{A_R(\ell_2)} \\
    &\notag \quad \times \int \frac{\dd[2]{q}}{(2\pi)^2} T(\ellvec'_1, -\ellvec'_1+\qvec, \ellvec'_2, -\ellvec'_2-\qvec)\, \tilde{W}(\qvec) \, \tilde{W}(-\qvec)\;. 
\end{align}
Considering only $\ell$ at small spatial scales well within the survey area and ignoring the impact of masks, $\Tilde{W}(\qvec)$ gives significant contributions only for $q \ll \ell$, and therefore we can approximate $P(|\ellvec+\qvec|)\simeq P(\ell)$. Then, as $W$ is either one or zero, 
\begin{align}
    \int \frac{\dd[2]{q}}{(2\pi)^2} \Tilde{W}(\ellvec+\qvec)\,\Tilde{W}(\qvec) &= \int \dd[2]{\alpha} W^2(\alphavec)\, \E^{-\I\,\alphavec\cdot\ellvec
    }\\
    &= \int \dd[2]{\alpha} W(\alphavec)\, \E^{-\I\,\alphavec\cdot\ellvec} = \Tilde{W}(\ellvec)\;,
\end{align}
we find
\begin{align}
\label{eq: Covariance power spectrum NKA}
    &\notag C_{\hat{P}}(\ell_1, \ell_2)\\
    &\notag\simeq\frac{2}{A^2}\int_{A_R(\ell_1)}\frac{\dd[2]{\ell'_1}}{A_R(\ell_1)}\int_{A_R(\ell_2)}\frac{\dd[2]{\ell'_2}}{A_R(\ell_2)}\, P(\ell'_1)\, P(\ell'_2)\,\\
    &\notag \qquad \times \tilde{W}(\ellvec'_1+\ellvec'_2)\, \tilde{W}(-\ellvec'_1-\ellvec'_2)\\
    &\quad +\frac{1}{A^2}\int_{A_R(\ell_1)}\frac{\dd[2]{\ell'_1}}{A_R(\ell_1)}\int_{A_R(\ell_2)}\frac{\dd[2]{\ell'_2}}{A_R(\ell_2)} \\
    &\notag\qquad \times \int \frac{\dd[2]{q}}{(2\pi)^2} T(\ellvec'_1, -\ellvec'_1+\qvec, \ellvec'_2, -\ellvec'_2-\qvec)\, \tilde{W}(\qvec) \, \tilde{W}(-\qvec)\\
     &\notag=:2\int_{A_R(\ell_1)}\frac{\dd[2]{\ell'_1}}{A_R(\ell_1)}\int_{ A_R(\ell_2)}\frac{\dd[2]{\ell'_2}}{A_R(\ell_2)} P(\ell'_1)\, P(\ell'_2)\, G_A(\ellvec'_1+\ellvec'_2)\\
    &\notag\quad+\int_{A_R(\ell_1)}\frac{\dd[2]{\ell'_1}}{A_R(\ell_1)}\int_{A_R(\ell_2)}\frac{\dd[2]{\ell'_2}}{A_R(\ell_2)}\\
    &\notag \qquad \times \int \frac{\dd[2]{q}}{(2\pi)^2} T(\ellvec'_1, -\ellvec'_1+\qvec, \ellvec'_2, -\ellvec'_2-\qvec)\, G_A(\qvec) \;, 
\end{align}
where we introduced the geometry factor $G_A$, defined as
\begin{align}
\label{eq: definition GA}
    G_A(\qvec) &=  \frac{1}{A^2}\tilde{W}(\qvec)\, \tilde{W}(-\qvec) \\
    &\notag =  \frac{1}{A^2}\int \dd[2]{\alpha_1}\, \int \dd[2]{\alpha_2} W(\alphavec_1)\, W(\alphavec_2)\, \E^{-\I\qvec\cdot(\alphavec_1-\alphavec_2)}\;.
\end{align}
The geometry factor contains the full dependence of $C_{\hat{P}}$ on the survey area. 

However, $C_{\hat{P}}$ in Eq.~\eqref{eq: Covariance power spectrum NKA} is not the same as $C^\mathrm{lit}_{\hat{P}}$ in Eq.~\eqref{eq: Covariance literature}. To convert $C_{\hat{P}}$ to $C^\mathrm{lit}_{\hat{P}}$, we need to perform the `large-field approximation'. For this approximation, we note that $G_A$ is related to the function $E_A(\etavec)$, which for a point $\alphavec$ inside $A$ gives the probability that a point $\alphavec+\etavec$ is also inside $A$, and is given by \citep{Heydenreich2020, Linke2023a}
\begin{equation}
    \label{eq: Definition EA}
    E_A(\etavec)=\frac{1}{A} \int_A \dd[2]{\alpha} W(\alphavec+\etavec)\;.
\end{equation}
With $E_A$, $G_A$ is
\begin{equation}
    G_A(\qvec) = \frac{1}{A}\int \dd[2]{\alpha} E_A(\alphavec)\, \E^{\I\alphavec\cdot\qvec}\;.
\end{equation}
The large-field approximation now assumes that $A$ is infinitely large so that $E_A(\alphavec)$ is unity for all $\alphavec$. Then,
\begin{equation}
    G_A(\qvec) \rightarrow \frac{(2\pi)^2}{A}\, \dirac(\qvec)\;.
\end{equation}
We define the result of $C_{\hat{P}}$ under this approximation as $C^\infty_{\hat{P}}$ , given as
\begin{align}
\label{eq: Covariance power spectrum large field approximation}
    C^\infty_{\hat{P}}(\ell_1, \ell_2)&=\frac{1}{A}\int_{A_R(\ell_1)}\frac{\dd[2]{\ell'_1}}{A_R(\ell_1)}\int_{A_R(\ell_2)}\frac{\dd[2]{\ell'_2}}{A_R(\ell_2)}\\
    &\notag \quad \times \left[ 2P^2(\ell'_1)\, (2\pi)^2 \dirac(\ellvec'_1+\ellvec'_2)+ T(\ellvec'_1, -\ellvec'_1, \ellvec'_2, -\ellvec'_2)\right] \;. 
\end{align}
These are exactly the first two terms of $C_{\hat{P}}^\mathrm{lit}$. Consequently, the full covariance $C_{\hat{P}}$ can be written as the sum of $C_{\hat{P}}^\infty$ and the `finite-field terms' $C_{\hat{P}}^\mathrm{FF}$. The finite-field terms are given by
\begin{align}
    C_{\hat{P}}^\mathrm{FF}&\notag=2\int_{A_R(\ell_1)}\frac{\dd[2]{\ell'_1}}{A_R(\ell_1)}\int_{A_R(\ell_2)}\frac{\dd[2]{\ell'_2}}{A_R(\ell_2)}P(\ell'_1)\, P(-\ell'_2)\, \\
    & \quad \times \Bigg[ G_A(\ellvec'_1+\ellvec'_2) - \frac{(2\pi)^2}{A} \dirac(\ellvec'_1+\ellvec'_2) \Bigg]\\
    &\notag+\frac{1}{A}\int_{A_R(\ell_1)}\frac{\dd[2]{\ell'_1}}{A_R(\ell_1)}\int_{A_R(\ell_2)}\frac{\dd[2]{\ell'_2}}{A_R(\ell_2)}\\
    &\notag \quad \times\Bigg[ \frac{1}{A} \int \frac{\dd[2]{q}}{(2\pi)^2} T(\ellvec_1, -\ellvec_1+\qvec, \ellvec_2, -\ellvec_2-\qvec)\, \tilde{W}(\qvec) \, \tilde{W}(-\qvec) \\
    &\notag\qquad- T(\ellvec_1, -\ellvec_1, \ellvec_2, -\ellvec_2)\Bigg]\;.
\end{align}
\citet{Takada2013} showed that the trispectrum can be approximated by
\begin{equation}
\label{eq: trispectrum approximation}
    T(\ellvec_1, -\ellvec_1+\qvec, \ellvec_2, -\ellvec_2-\qvec) \simeq T(\ellvec_1, -\ellvec_1, \ellvec_2, -\ellvec_2) + T_\mathrm{SSC}(\ellvec_1, \ellvec_2, \qvec)\;,
\end{equation}
with the individual parts given as Limber integrals over the trispectrum expressions in their Equation (32). Notably, $T_\mathrm{SSC}$ is proportional to the linear matter power spectrum $P_L(\qvec/\chi)$ at wavevector $\qvec/\chi$ and comoving distance $\chi$. Therefore, $T_\mathrm{SSC}$ vanishes for $|\qvec|=0$\;.
With Eq.~\eqref{eq: trispectrum approximation}, one obtains
\begin{align}
\label{eq: C_P-C_Pinf}
    \notag C_{\hat{P}}^\mathrm{FF}&\simeq\frac{2}{A}\int_{A_R(\ell_1)}\frac{\dd[2]{\ell'_1}}{A_R(\ell_1)}\int_{A_R(\ell_2)}\frac{\dd[2]{\ell'_2}}{A_R(\ell_2)} P(\ell'_1)\,P(\ell'_2)\,\\
    & \quad \times \Big[\frac{1}{A}\tilde{W}(\ellvec'_1+\ellvec'_2)\, \tilde{W}(-\ellvec'_1-\ellvec'_2)-(2\pi)^2\,\dirac(\ellvec'_1+\ellvec'_2)\Big]\\
        &\notag+\frac{1}{A^2}\int_{A_R(\ell_1)}\frac{\dd[2]{\ell'_1}}{A_R(\ell_1)}\int_{A_R(\ell_2)}\frac{\dd[2]{\ell'_2}}{A_R(\ell_2)}\\
        &\notag \quad \times\int\frac{\dd[2]{q}}{(2\pi)^2} T_\mathrm{SSC}(\ellvec_1, \ellvec_2, \qvec)\, \tilde{W}(\qvec) \, \tilde{W}(-\qvec)\\
    &\notag=C_{\hat{P}}^\mathrm{FF, G}+C_{\hat{P}}^\mathrm{SSC}\;. 
\end{align}
The second summand is the SSC, while the first summand is a Gaussian term, which is usually negative and suppresses the Gaussian covariance. As becomes apparent in our derivation, the SSC and the Gaussian suppression term can both be seen as corrections for the large-field approximation, which is why we collectively refer to them as finite-field terms. We note that $C_{\hat{P}}^\infty(\ell_1, \ell_2)$ only depends on the power spectrum and trispectrum at $\ell$ close to $\ell_1$ and $\ell_2$, because the integrals in Eq.~\eqref{eq: Covariance power spectrum large field approximation} only go over the $A_R(\ell_i)$. In contrast, the SSC depends on the trispectrum at all $\ell$ modes because the integral over the $\qvec$ extends over all $\realspace^2$. Therefore, for the power spectrum, the SSC captures the dependence of the covariance on modes outside the survey area. We will see in the following sections that this intuition does not hold for real space statistics. For this, in the following section we derive the covariance for an estimator of $\Xi$ in real space.

\section{\label{sec: xi real space} Covariance of a real space statistic}

We now consider a localised second-order shear statistic $\Xi$ in real space that can be written as
\begin{align}
\label{eq: definition second order statistic}
    &\Xi(\theta, \alphavec) \\
    &\notag = \int \dd[2]{\vartheta_1} \int \dd[2]{\vartheta_2} U_1(\theta, \varthetavec_1-\alphavec)\, U_2(\theta, \varthetavec_2-\alphavec)\, \expval{\kappa(\varthetavec_1)\, \kappa(\varthetavec_2)}\;,
\end{align}
where the $U_a$ are two filter functions of scale $\theta$, and $\kappa$ is the weak-lensing convergence. Examples of such a statistic are the COSEBIs \citep{Schneider2010} or the second-order aperture statistics $\MapMap$ \citep{Schneider1998}. We note that while we are writing the statistic here in terms of the (unobservable) convergence for simplicity, $\Xi$ can also be written in terms of the weak lensing shear $\gamma$, which is observable if $U_1$ and $U_2$ are compensated filter functions. This is the case for both COSEBIs and $\MapMap$.

\begin{figure}
\centering
    \begin{tikzpicture}
    \draw (0,0) rectangle (5,5);
    \draw (5,0) node[anchor=south east]{$A'$};
    \draw [fill=pink](1,1) rectangle (4,4);
    \draw (4,1) node[anchor=south east]{$A$};
    
    \filldraw [black] (1.5, 2) circle (2pt);
    \draw (1.5, 2) node[anchor=west]{$\alphavec$};
    \draw (1.5, 2) circle (1);
    \draw (1.5, 2) -- (0.5,2);
    \draw (0.8, 2) node[anchor=north]{$\theta$};

    \filldraw [black] (2.5, 4.5) circle (2pt);
    \draw (2.5, 4.5) node[anchor=west]{$\alphavec'$};
    \draw (2.5, 4.5) circle (1);
    \draw (2.5, 4.5) -- (2.5, 3.5);
    \draw (2.5, 3.75) node[anchor=west]{$\theta$};
    \end{tikzpicture}
\caption{Illustration of the estimation of the statistic $\Xi$. The area $A'$ is the size of the full convergence field, which we convolve with filter functions of scale radius $\theta$, illustrated by the circles. The convolution results at positions $\alphavec$ within the smaller area $A$ only depend on $\kappa$ within $A'$, while for positions $\alphavec'$ outside of $A$, information on $\kappa$ outside of $A'$ is needed for an unbiased estimate. Therefore, the border of $A'$ outside of $A$ is discarded before obtaining $\Xi$.}
\label{fig: estimation}
\end{figure}

To estimate the statistic from a convergence field $\kappa$ of size $A'$, we can convolve $\kappa$ with the filter functions and then average over the pixel values. However, due to the finiteness of the survey area, the convolution result at the survey boundaries is biased. Therefore, the average needs to be taken over a smaller area $A$, which excludes the borders of the field (see Fig.~\ref{fig: estimation}). This leads to the estimator
\begin{align}
    \hat{\Xi}(\theta) &= \frac{1}{A}\int_A \dd[2]{\alpha} \prod_{i=1}^{2} \int_{A'}\dd[2]{\vartheta_i} U_i(\theta,\varthetavec_i-\alphavec)\,\kappa_i\;,
\end{align}
where $\kappa_i=\kappa(\varthetavec_i)$. Under the assumption that $U_i(\theta,\varthetavec_i-\alphavec)$ vanishes for $\varthetavec_i$ outside of $A'$ for all $\alphavec \in A$, we can replace the integral over $A'$ by an integral over the whole $\realspace^2$. With this, and the survey window function $W(\varthetavec)$, which is one for $\varthetavec$ inside $A$ and zero otherwise,
\begin{align}
\label{eq: Definition Estimator}
    \hat{\Xi}(\theta) &= \frac{1}{A}\int \dd[2]{\alpha} W(\alphavec)\,\prod_{i=1}^{2} \int \dd[2]{\vartheta_i} U_i(\theta,\varthetavec_i-\alphavec)\,\kappa_i\;.
\end{align}
The covariance of $\hat{\Xi}$ is 
\begin{equation}
    C_{\hat{\Xi}}(\theta_1, \theta_2) = \expval{\hat{\Xi}(\theta_1)\,\hat{\Xi}(\theta_2)} - \expval{\hat{\Xi}(\theta_1)}\,\expval{\hat{\Xi}(\theta_2)}\;.
\end{equation}
With Eq.~\eqref{eq: Definition Estimator},
\begin{align}
    \label{eq: covariance zeroth step}\expval{\hat{\Xi}(\theta_1)\,\hat{\Xi}(\theta_2)} &= \frac{1}{A^2}\int \dd[2]{\alpha_1} \int \dd[2]{\alpha_2} W_A(\alphavec_1)\, W_A(\alphavec_2)
    \\&\notag \quad \times 
    \int \dd[2]{\vartheta_1} \int \dd[2]{\vartheta_2} \int \dd[2]{\vartheta_3} \int \dd[2]{\vartheta_4}     \expval{\kappa_1\, \kappa_2\, \kappa_3\, \kappa_4}\\
    &\notag \quad \times U_1(\theta_1, \varthetavec_1-\alphavec_1)\,U_2(\theta_1, \varthetavec_2-\alphavec_1)
    \\ &\notag \quad \times 
    U_1(\theta_2, \varthetavec_3-\alphavec_2)\,U_2(\theta_2, \varthetavec_4-\alphavec_2)
\\
    &\label{eq: covariance first step}= \frac{1}{A^2}\int \dd[2]{\alpha_1}\int \dd[2]{\alpha_2} W_A(\alphavec_1)\, W_A(\alphavec_2)
    \\ &\notag \quad \times
\expval{\varkappa_1(\theta_1,\alphavec_1)\,\varkappa_2(\theta_1,\alphavec_1)\,\varkappa_1(\theta_2,\alphavec_2)\,\varkappa_2(\theta_2,\alphavec_2)}\;,
\end{align}
where we introduced the smoothed convergence field $\varkappa_a$,
\begin{equation}
    \varkappa_a(\theta,\alphavec) = \int \dd[2]{\varthetavec} U_a(\theta,\varthetavec-\alphavec)\, \kappa(\varthetavec)\;.
\end{equation}
For ease of notation we define the field $\varkappa^2(\theta_1,\alphavec):=\varkappa_1(\theta_1,\alphavec)\,\varkappa_2(\theta_1,\alphavec)$. The expectation value in Eq.~\eqref{eq: covariance first step} is a second-order correlation function $\xi_{\varkappa^2}$ of this field and is defined as
\begin{align}
    \xi_{\varkappa^2}(\theta_1, \theta_2, \etavec) &= \expval{\varkappa^2(\theta_1, \alphavec)\,\varkappa^2(\theta_2, \alphavec+\etavec)} \\
    &\notag = \expval{\varkappa_1(\theta_1, \alphavec)\,\varkappa_2(\theta_1, \alphavec)\,\varkappa_1(\theta_2, \alphavec+\etavec)\,\varkappa_2(\theta_2, \alphavec+\etavec)} \;.
\end{align}
With $\xi_{\varkappa^2}$ and the function $E_A$, defined in Eq.~\eqref{eq: Definition EA},
\begin{align}
\label{eq: covariance from correlation functions}
    C_{\hat{\Xi}}(\theta_1, \theta_2) &= \frac{1}{A^2}\, \int \dd[2]{\alpha_1} \int \dd[2]{\alpha_2} W_A(\alphavec_1)\, W_A(\alphavec_2)\,
    \\    &\notag \quad \times
    \xi_{\varkappa^2}(\theta_1,\theta_2,|\alphavec_1-\alphavec_2|) -  \expval{\hat{\Xi}(\theta_1)}\,\expval{\hat{\Xi}(\theta_2)} \\
    &\notag= \frac{1}{A}\int \dd[2]{\eta} E_A(\etavec)\, \xi_{\varkappa^2}(\theta_1,\theta_2,\etavec) - \expval{\hat{\Xi}(\theta_1)}\,\expval{\hat{\Xi}(\theta_2)}\;.
\end{align}
In this expression, the covariance of $\Xi$ can be inferred from a two-point correlation function $\xi_{\varkappa^2}$. Notably, $\xi_{\varkappa^2}$ needs to be known only for $\etavec$ inside the survey area, as $E_A$ vanishes outside. Consequently, $C_{\hat{\Xi}}$ does not depend on any information on spatial scales larger than the survey area. 

This finding might appear to contradict Eq.~\eqref{eq: Covariance power spectrum NKA}, which shows that the power spectrum covariance (in particular, the SSC) clearly depends on Fourier modes larger than the survey. However, Eq.~\eqref{eq: covariance from correlation functions} takes a `real space' view. The real-space correlation function $\xi_{\varkappa^2}$ at spatial scales within the survey window is impacted by the power spectrum and trispectrum at Fourier modes larger than the survey window. Therefore, while the covariance estimation in Fourier space requires all modes, in real space, we can limit ourselves to spatial scales within the survey window. Thus, while information from `super-survey' Fourier modes is needed, no information on `super-survey' spatial scales is required.

\section{Covariance estimation from correlation functions}
\label{sec: cov corr}

An interesting aspect of Eq.~\eqref{eq: covariance from correlation functions} is that it allows the covariance of $\hat{\Xi}$ to be quickly evaluated for various survey geometries. Given a single, full-sky simulation of the convergence field, $\xi_{\varkappa^2}$ and $\expval{\hat{\Xi}(\theta_1)}$ can be measured with high accuracy. Then, $ C_{\hat{\Xi}}$ can be estimated for any survey geometry, simply by adapting the $E_A(\etavec)$ in Eq.~\eqref{eq: covariance from correlation functions}.

Another application of Eq.~\eqref{eq: covariance from correlation functions} is to estimate $\xi_{\varkappa^2}$ as the average of $\xi_{\varkappa^2}^{(i)}$ measured for each realisation $i$ of an ensemble of simulations. This approach delivers a covariance estimate that coincides with the standard sample covariance estimate. To see this, we use 
\begin{equation}
\xi^{(i)}_{\varkappa^2}(\theta_1, \theta_2, \etavec) = \frac{1}{A}\frac{1}{E_A(\etavec)}\,\int_A \dd[2]{\alphavec} {\varkappa^2_{(i)}}(\alphavec, \theta_1)\,{\varkappa^2_{(i)}}(\alphavec+\etavec, \theta_2)\, W(\alphavec+\etavec)\;,
\end{equation}
and 
\begin{equation}
  \xi_{\varkappa^2}(\theta_1, \theta_2, \etavec) = \frac{1}{N} \sum_{i=1}^N   \xi^{(i)}_{\varkappa^2}(\theta_1, \theta_2, \etavec)\;,
\end{equation}
where $N$ is the number of realisations. We further use
\begin{align}
    \expval{\hat{\Xi}(\theta)} 
    &= \frac{1}{A}\frac{1}{N}\sum_{i=1}^N \int \dd[2]{\alpha} W_A(\alphavec)\, {\varkappa^2_{(i)}}(\alphavec, \theta)\\
    &=: \frac{1}{N}\sum_{i=1}^N \hat{\Xi}^{(i)}(\theta)\;.
\end{align}
Then, with Eq.~\eqref{eq: covariance from correlation functions}, the covariance ${C}^\mathrm{corr}_{\hat{\Xi}}$ from the correlation functions is
\begin{align}
    {C}^\mathrm{corr}_{\hat{\Xi}}(\theta_1, \theta_2) &= \frac{1}{N\,A^2}\sum_{i=1}^N \int \dd[2]{\eta} \int \dd[2]{\alpha} W_A(\alphavec)\, W_A(\alphavec+\etavec)\,\\
    &\notag \qquad\times \varkappa^2_{(i)}(\alphavec, \theta_1)\, \varkappa^2_{(i)}(\alphavec+\etavec, \theta_2) \\
    &\notag \quad -  \frac{1}{N^2} \sum_{i=1}^N \sum_{j=1}^N \hat{\Xi}^{(i)}(\theta_1)\, \hat{\Xi}^{(j)}(\theta_2)\\
    \notag&= \frac{1}{N} \sum_{i=1}^N \hat{\Xi}^{(i)}(\theta_1)\, \hat{\Xi}^{(i)}(\theta_2)  -  \frac{1}{N^2} \sum_{i=1}^N \sum_{j=1}^N \hat{\Xi}^{(i)}(\theta_1)\, \hat{\Xi}^{(j)}(\theta_2)\\
    &= \frac{N-1}{N}\, C^\mathrm{sim}_{\hat{\Xi}}(\theta_1, \theta_2)\;,
\end{align}
with the sample covariance  $C^\mathrm{sim}_{\hat{\Xi}}$ defined as
\begin{equation}
     C^\mathrm{sim}_{\hat{\Xi}} = \frac{1}{N-1} \sum_{i=1}^N \left(\hat{\Xi}^{(i)}(\theta_1)- \expval{\hat{\Xi}(\theta_1)} \right)\, \left(\hat{\Xi}^{(i)}(\theta_2)- \expval{\hat{\Xi}(\theta_2)} \right)\;.
\end{equation}
Consequently, estimating the covariance from the average $\varkappa^2$ correlation function of an ensemble of simulations is equivalent to estimating it from the $\Xi$ measured individually on each realisation.

 We validate this finding by specifying our statistic in the remainder of this section to the second-order aperture mass $\MapMap(\theta)$ with the filter function by \citet{Crittenden2002}:
\begin{equation}
    U_1(\theta, \varthetavec) =  U_2(\theta, \varthetavec) = U(\theta, \vartheta)=\frac{1}{2\pi\theta^2} \, \left(1-\frac{\vartheta^2}{2\theta^2}\right)\, \exp(-\frac{\vartheta^2}{2\theta^2})\;.
\end{equation}
Then, setting $\varkappa^2=\Map^2$, which is given as
\begin{align}
\label{eq: definition Map2}
    &\Map^2(\theta; \varthetavec) \\
    &\notag= \int \dd[2]{\alpha_1} \int \dd[2]{\alpha_2} U(\theta,|\varthetavec-\alphavec_1|)\, U(\theta, |\varthetavec-\alphavec_2|)\,\kappa(\alphavec_1)\,\kappa(\alphavec_2)\,,
\end{align}
and $\xi_{\varkappa^2}$ becomes
\begin{equation}
    \xi_{\Map^2}(\theta_1, \theta_2, \etavec) = \expval{ \Map^2(\theta_1; \varthetavec)\,  \Map^2(\theta_2; \varthetavec+\etavec)}\;.
\end{equation}
This expression can be generalised to higher orders of $\Map^n$ with $n>2$. \citet{Porth2021} give the expressions for the variance of $\Map^n$ estimated with the so-called direct estimator as a function of the correlation functions of $\Map^n$ for a single aperture.
We validate Eq.~\eqref{eq: covariance from correlation functions} for the aperture statistics by measuring the covariance $C_{\MapMapEst}$ of $\MapMapEst$ in convergence maps from the Scinet LIghtcone Simulations \citep[SLICS,][]{HarnoisDeraps2015}, the details of which are described in Appendix~\ref{app: validation: data}. We estimate the covariance of $\MapMap$ using the sample covariance and correlation function-based approach(see Appendix~\ref{app: validation: measurement}).

\begin{figure*}
    \centering
    \includegraphics[width=\linewidth]{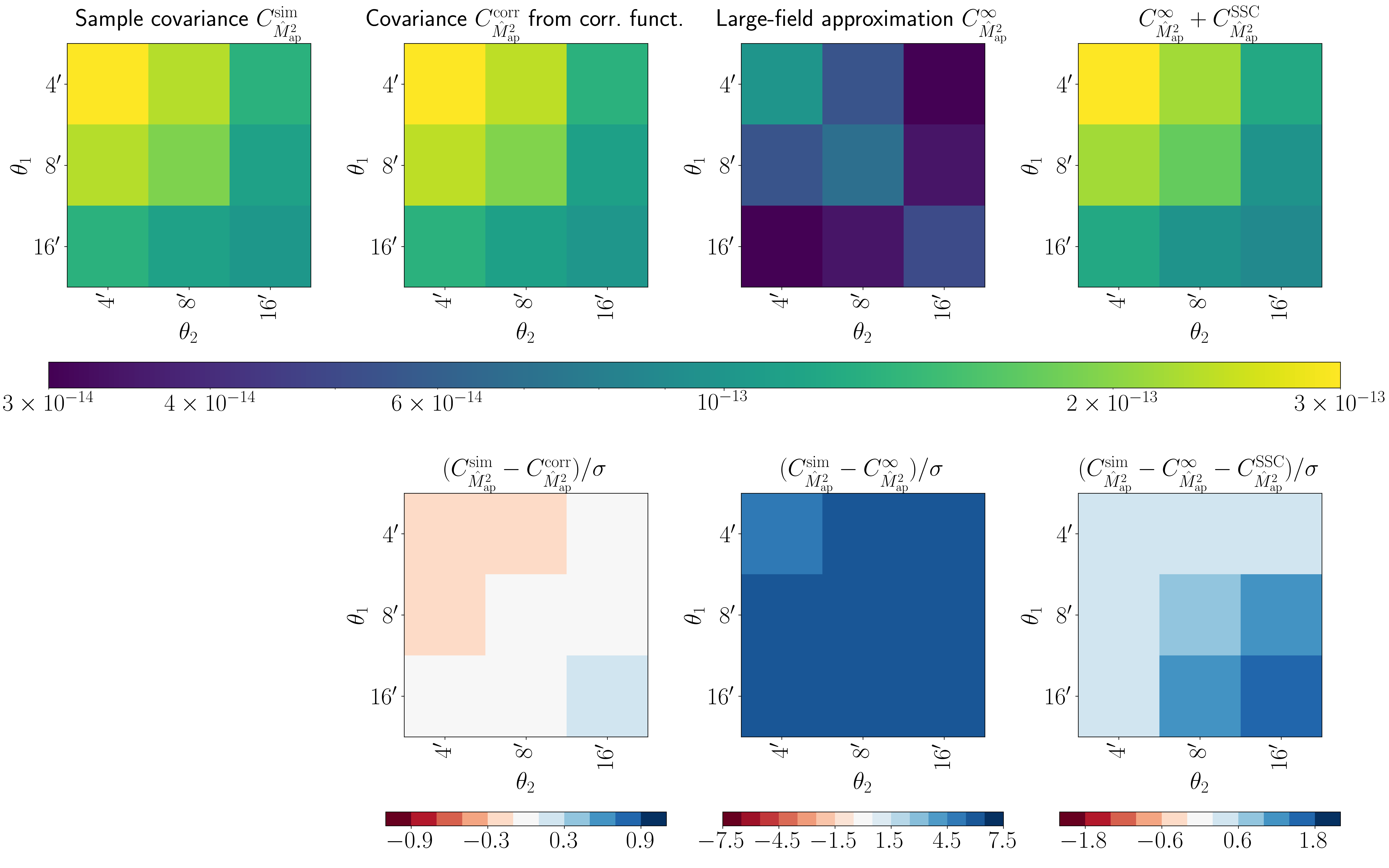}
    \caption{Comparison of $\MapMapEst$ covariance obtained in different ways. Upper row, from left to right: Sample covariance from the SLICS, covariance estimated from $\Map$ correlation functions in the SLICS, model covariance under the large-field approximation, full model covariance including the finite field terms. Lower row: Differences between sample covariance and other covariance estimates, normalised by bootstrap uncertainty of the sample covariance. We note that the colour bar changes between the three bottom plots.}
    \label{fig: all Cov estimates}
\end{figure*}

The covariance estimates and their difference are shown in the first two panels of Fig.~\ref{fig: all Cov estimates}. The two estimates almost coincide. Accordingly, the correlation-function-based approach captures the full covariance, even though $\xi_{\Map^2}$ is known only for scales within the survey area. Consequently, as expected from Eq.~\eqref{eq: covariance from correlation functions}, no information on spatial scales outside the survey is needed for an accurate covariance estimate.

\section{\label{sec: xi fourier space} Connection between real and Fourier space statistics}

The statistic $\Xi$ defined in Eq.~\eqref{eq: definition second order statistic} can be expressed in terms of the power spectrum as
\begin{equation}
    \Xi (\theta) = \int \frac{\dd[2]{\ell}}{(2\pi)^2} \tilde{U}_1(\theta, \ellvec)\,\tilde{U}_2(\theta, \ellvec)\,P(\ell)\;,
\end{equation}
where the $\Tilde{U}_i$ are the Fourier transforms of the filter functions $U_i$. A common strategy \citep[e.g.,][]{Joachimi2021, Friedrich2021} to model the covariance of $\Xi$
is to use
\begin{align}
    \label{eq: relation cov Xi cov P naive}
    C_{\Xi}(\theta_1, \theta_2) &= \int \frac{\dd[2]{\ell_1}}{(2\pi)^2} \int \frac{\dd[2]{\ell_2}}{(2\pi)^2} \tilde{U}_1(\theta_1, \ellvec_1)\,\tilde{U}_2(\theta_1, \ellvec_1)\\
    &\notag \times \tilde{U}_1(\theta_2, \ellvec_2)\,\tilde{U}_2(\theta_2, \ellvec_2)\, C_{\hat{P}}(\ell_1, \ell_2)\;.
\end{align}
However, this approach is not necessarily correct. To show this, we relate $C_{\hat{\Xi}}$ to the power- and trispectrum to compare it to previous expressions of second-order shear covariances and to discuss the SSC for real-space statistics. As shown in Appendix~\ref{app: covariance calculation}, $ C_{\hat{\Xi}}(\theta_1, \theta_2)$ can be expressed as
\begin{align}
    C_{\hat{\Xi}}(\theta_1, \theta_2) 
    &=\label{eq: covariance complete} \int \frac{\dd[2]{\ell}_1}{(2\pi)^2} \int \frac{\dd[2]{\ell}_2}{(2\pi)^2}\, G_A(\ellvec_1+\ellvec_2)\,  P(\ell_1)\,P(\ell_2)\,
    \\    &\notag \quad \times 
    \left(\tilde{U}_1(\theta_1, \ellvec_1)\, \tilde{U}_1(\theta_2, \ellvec_1)\,\tilde{U}_2(\theta_1, \ellvec_2)\, \tilde{U}_2(\theta_2, \ellvec_2)
    \right.\\    &\notag \qquad \left.
    +\tilde{U}_1(\theta_1, \ellvec_1)\, \tilde{U}_1(\theta_2, \ellvec_2)\,\tilde{U}_2(\theta_1, \ellvec_2)\, \tilde{U}_2(\theta_2, \ellvec_1)\right) \\
    &\notag \quad +\int \frac{\dd[2]{\ell}_1}{(2\pi)^2} \int \frac{\dd[2]{\ell}_2}{(2\pi)^2} \int \frac{\dd[2]{\ell}_3}{(2\pi)^2}\,  G_A(\ellvec_1+\ellvec_2)\,
    \\    &\notag \quad \times 
    T(\ellvec_1, \ellvec_2, \ellvec_3, -\ellvec_1-\ellvec_2-\ellvec_3)\,  \tilde{U}_1(\theta_1, \ellvec_1)\,\tilde{U}_2(\theta_1, \ellvec_2) \\
    &\notag \quad \times \tilde{U}_1(\theta_2, \ellvec_3)\, \tilde{U}_2(\theta_2, \ellvec_1+\ellvec_2+\ellvec_3)\;.
\end{align}
One notices a Gaussian and a non-Gaussian part of the covariance, with the Gaussian part depending on the power spectrum and the non-Gaussian part depending on the trispectrum, similar to the exact covariance of the power spectrum in Eq.~\eqref{eq: Covariance power spectrum unapproximated}. However, $C_{\hat{\Xi}}$ is not in the form of Eq.~\eqref{eq: relation cov Xi cov P naive} which is a weighted integral over $C_{\hat{P}}$. We also notice three differences when directly comparing Eq.~\eqref{eq: covariance complete} to Equations (E1) and (E7) in \citet{Joachimi2021}. First, neither the Gaussian term nor the non-Gaussian term scales with the inverse of the survey area $A$, and instead both show a more complicated dependence on survey geometry via $G_A$. Second, the non-Gaussian term depends on the trispectrum for all $\ell$-configurations, not simply for parallelograms with $\ellvec_3=\ellvec_1$. Third, the SSC does not appear as an additional term.

To reconcile Eq.~\eqref{eq: covariance complete} with the expressions in \citet{Joachimi2021}, we need to perform the large-field approximation. As mentioned in Sect.~\ref{sec: power spectrum}, under this approximation,  $G_A$ is proportional to a Dirac delta, so the covariance becomes
\begin{align}
\label{eq: covariance large-field approximation}
    &C^\infty_{\hat{\Xi}}(\theta_1, \theta_2)\\
    &\notag=\frac{2}{A}\,\int \frac{\dd[2]{\ell}}{(2\pi)^2}\,  P^2(\ell)\, \tilde{U}_1(\theta_1, \ellvec)\,\tilde{U}_1(\theta_2, \ellvec)\, \tilde{U}_2(\theta_1, \ellvec)\, \tilde{U}_2(\theta_2, \ellvec)\\
    &\notag \quad +\frac{1}{A}\int \frac{\dd[2]{\ell}_1}{(2\pi)^2} \int \frac{\dd[2]{\ell}_2}{(2\pi)^2} \, T(\ellvec_1, -\ellvec_1, \ellvec_2, -\ellvec_2)\,\\
    &\notag \quad\quad \times    \tilde{U}_1(\theta_1, \ellvec_1)\,\tilde{U}_2(\theta_1, \ellvec_1)\, \tilde{U}_1(\theta_2, \ellvec_2)\, \tilde{U}_2(\theta_2, \ellvec_2)\;.
\end{align}
This is equivalent to the commonly used expressions for the Gaussian and intra-survey non-Gaussian covariance for a second-order statistic (Equations E1 and E7 in \citealp{Joachimi2021}). In particular, we recover the scaling with the inverse survey area and the dependence on parallelogram $\ell$-configurations for the non-Gaussian part. Consequently, the sum of the Gaussian and intra-survey non-Gaussian covariance can be considered as an approximation of the exact covariance in Eq.~\eqref{eq: covariance complete} for very large survey windows.

A comparison of Eq.~\eqref{eq: covariance large-field approximation} to Eq.~\eqref{eq: Covariance power spectrum large field approximation} shows that $C^\infty_{\hat{\Xi}}$ is given by the integral over the large-field approximation of the power spectrum covariance $C^\infty_{\hat{P}}$. Therefore, the approach to obtain the covariance of the real-space statistic from the power spectrum covariance is correct if the large-field approximation holds. However, the large-field approximation neglects a significant part of the covariance. We can see this for $\MapMap$, for which we calculate Eq.~\eqref{eq: covariance large-field approximation} according to Appendix~\ref{app: validation: modelling}. The third panel of Fig.~\ref{fig: all Cov estimates} shows the covariance for $\MapMap$, which is modelled with Eq.~\eqref{eq: covariance large-field approximation}, along with the fractional difference from the sample covariance estimate from the SLICS (see Appendix~\ref{app: validation}). The approximation $C_{\MapMapEst}^\infty$ is significantly too small, with deviations of more than five times the statistical uncertainty on the sample covariance. 
This large difference is not surprising given that the survey area $A$ here is only $62\, \mathrm{deg}^2$. However, even for the Kilo-degree survey (KIDS) data release KiDS-1000, $C^\infty$ cannot describe the covariance of the cosmic shear band powers \citep{Joachimi2021}.

In analogy to the power spectrum case, we define the difference between $C_{\hat{\Xi}}$ and $C_{\hat{\Xi}}^\infty$ as finite-field terms $C_{\hat{\Xi}}^\mathrm{FF}$ for $\hat{\Xi}$. To calculate this term, we can perform the same approximation as that performed for the power spectrum covariance, namely that the modes $\ell, \ell_1,$ and $\ell_2$ are large compared to the modes $q$ on which $G_A$ varies. With this,
\begin{align}
\label{eq: covariance with ssc}
    &\notag C^\mathrm{FF}_{\hat{\Xi}}(\theta_1, \theta_2)\\
    &\notag=  \int \frac{\dd[2]{\ell}_1}{(2\pi)^2} \int \frac{\dd[2]{\ell}_2}{(2\pi)^2} \tilde{U}_1(\theta_1, \ellvec_1)\,\tilde{U}_2(\theta_1, \ellvec_1)\, \tilde{U}_1(\theta_2, \ellvec_2)\, \tilde{U}_2(\theta_2, \ellvec_2)\\
    &\quad \times \Bigg\lbrace P(\ell_1)\,P(\ell_2)\,\left[G_A(\ellvec_1+\ellvec_2) - \frac{(2\pi)^2}{A}\dirac(\ellvec_1+\ellvec_2)\right]\\
   &\notag\qquad + \int \frac{\dd[2]{q}}{(2\pi)^2}  G_A(\qvec)\,
    T_\mathrm{SSC}(\ellvec_1, \ellvec_2, \qvec)\Bigg\rbrace\\
    &= \notag C^\mathrm{FF, G}_{\hat{\Xi}}(\theta_1, \theta_2) + C^\mathrm{SSC}_{\hat{\Xi}}(\theta_1, \theta_2)\;,
\end{align}
where $T_\mathrm{SSC}$ is the same as in Eq.~\eqref{eq: trispectrum approximation}. To show that $C_{\hat{\Xi}}$ is indeed composed of $C_{\hat{\Xi}}^\infty$ and $C_{\hat{\Xi}}^\mathrm{FF}$, we show in Fig.~\ref{fig: all Cov estimates} the modelled covariance of $\MapMap$ including the finite-field terms and the sample covariance estimate in the SLICS. We see that the covariances coincide within the bootstrap uncertainties of the sample covariance. 

\begin{figure*}
    \centering
    \includegraphics[width=\linewidth]{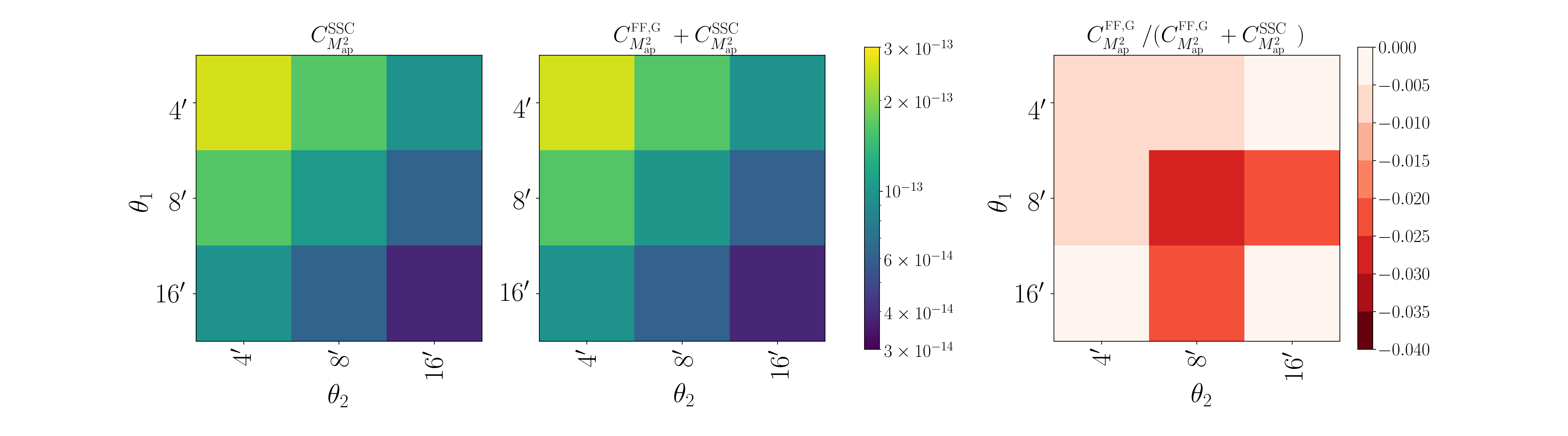}
    \caption{Comparison of finite field terms for $\MapMapEst$ for the SLICS setup. Left:   $C^\mathrm{SSC}_{\MapMapEst}$, which is equivalent to Equation (E10) in \citet{Joachimi2021}. Centre: Both finite-field terms. Right: Ratio of $C^\mathrm{FF, G}_{\MapMapEst}$ to complete $C^\mathrm{FF}_{\MapMapEst}$.}
    \label{fig: ssc}
\end{figure*}

The first summand $C^\mathrm{FF, G}_{\hat{\Xi}, 1}$ in  Eq.~\eqref{eq: covariance with ssc} depends on the power spectrum and therefore already occurs for Gaussian fields. In general, it is negative and decreases the magnitude of the other Gaussian covariance term. This effect has already been noticed for the covariance of shear correlation functions in \citet{Sato2011}. 

In \citet{Joachimi2021}, only the second summand  $C^\mathrm{SSC}_{\hat{\Xi}}$ in Eq.~\eqref{eq: covariance with ssc} is considered (see their Equation E10), while the first summand $C^\mathrm{FF}_{\hat{\Xi}}$ is neglected. However, at least for the second-order aperture statistics, this neglect has only a small impact, because the $ C^\mathrm{FF, G}_{\MapMapEst}$ is small compared to $ C^\mathrm{SSC}_{\MapMapEst}$. This can be seen in Fig.~\ref{fig: ssc}, where we compare the full finite-field term to $ C^\mathrm{SSC}_{\MapMapEst}$ for the $\MapMapEst$ covariance in the SLICS. The first finite-field term accounts for less than 5\% of the total $C^\mathrm{FF}_{\MapMapEst}$. Therefore, using just the SSC is accurate enough to describe the sample covariance in the SLICS.

It is important to note here that both $C_{\hat{\Xi}}^\infty$ and $C_{\hat{\Xi}}^\mathrm{FF}$ depend on the power spectrum and trispectrum throughout the $\ell$-space. Consequently, modes larger than the survey area impact not only the SSC term but also $C_{\hat{\Xi}}^\infty$. The common interpretation that the SSC captures the full impact of `super-survey' $\ell$-modes is incorrect for real-space statistics.

\section{\label{sec: conclusion} Conclusion}
We derived the full covariance $C_{\hat{\Xi}}$ for a localised, second-order statistic $\Xi$ in real space and compared it to the covariance $C_{\hat{P}}$ of the lensing power spectrum. Both covariances depend on the exact survey geometry. Under the large-field approximation, which is the limit for a broad window function, these covariances reduce to approximated terms $C_{\hat{\Xi}}^\infty$ and $C_{\hat{P}}^\infty$, which scale with the inverse survey area. While we define $\Xi$ in terms of the convergence $\kappa$, we note that for compensated filter functions, $\Xi$ can be equivalently written in terms of the weak lensing shear, so all our conclusions are valid for such shear statistics as well.

We find that the difference between $C_{\hat{P}}$ and $C_{\hat{P}}^\infty$ gives rise to two terms, which we collectively refer to as finite-field terms $C_{\hat{P}}^\mathrm{FF}$. While $C_{\hat{P}}^\infty$ scales linearly with inverse survey area, the finite-field terms show a complicated dependence on the survey geometry. One of these terms is the SSC $C_{\hat{P}}^\mathrm{SSC}$, which, in contrast to the other terms, depends on the matter trispectrum on all $\ell$-modes including those larger than the survey area.

The covariance $C_{\hat{\Xi}}$ can also be written as the sum of a large-field approximation $C_{\hat{\Xi}}^\infty$ and finite-field terms $C_{\hat{\Xi}}^\mathrm{FF}$. However, both $C_{\hat{\Xi}}^\infty$ and $C_{\hat{\Xi}}^\mathrm{FF}$ depend on the power spectrum and trispectrum at all $\ell$-modes, including those larger than the survey area. Therefore, the label `super-sample' is slightly misleading for the SSC of a real-space statistic, as it is not the only term containing super-sample information. 

The second finite field term in addition to the SSC depends on the power spectrum and is already present in Gaussian fields. This term essentially decreases the Gaussian covariance, an effect already noted for shear correlation functions by \citet{Sato2011}. While we show here that neglecting this term \citep[e.g.][]{Joachimi2021} is accurate for the second-order aperture statistics $\MapMapEst$, \citet{Shirasaki2019} and \citet{Troxel2018} found that for the shear correlation functions $\xi_+$ and $\xi_-$ ignoring this effect leads to a significant overestimation of the covariance. We suspect this difference occurs because the aperture mass filter function is compensated. Therefore, the average aperture mass vanishes and the integral in the Gaussian finite field term is essentially `cut-off' at small $\ell$. However, for the correlation functions, small $\ell$ need to be taken into account. 

We show that the covariance $C_{\hat{\Xi}}$ of the real-space statistic cannot be obtained from the power spectrum covariance without the large-field approximation. The commonly used transformation between power spectrum covariance and real-space covariance only holds for $C_{\hat{P}}^\infty$ and $C_{\hat{\Xi}}^\infty$. This finding is not surprising. A linear transform between the $C_{\hat{P}}$ and $C_{\hat{\Xi}}$ is mathematically only possible if the estimators $\hat{P}$ and $\hat{\Xi}$ are related linearly. While $P$ and $\Xi$ are indeed related by a Fourier transform, the estimators are generally not. Consequently, one would not expect to simply transform one covariance into the other.
    
Finally, we demonstrate that $C_{\hat{\Xi}}$ can be fully determined from correlation functions of smoothed convergence maps known only at spatial scales smaller than the survey area. We show that if one estimates the correlation functions from an ensemble of simulations, this approach gives the same result (up to a prefactor) as estimating the statistics directly on each realisation and taking the sample covariance. This indicates that correlations outside the survey area do not influence the covariance of a second-order shear statistic in real space.

This finding is not surprising given that \citet{Schneider2002} already showed that the Gaussian covariance of shear correlation functions is given by second-order correlation functions of galaxy ellipticities known inside the survey area. However, it provides an alternative method to estimate covariances for varying survey geometries quickly. If the $\varkappa^2$-correlation function is known accurately, for example from a single, high-precision simulation, the covariance of the statistics $\hat{\Xi}$ can be calculated by specifying the survey geometry function $E_A$ and evaluating Eq.~\eqref{eq: covariance from correlation functions}. Consequently, for this case, changing the survey geometry would not require new simulations to be produced for covariance estimates, provided the redshift distribution of source galaxies remains the same.

While this paper is concerned with second-order statistics only, finite-field terms also play a significant role for higher-order statistics. \citet{Linke2023a} showed that for third-order aperture statistics $\MapMapMap$, several finite-field terms occur, with one of them dominating the Gaussian covariance. In contrast, for the convergence probability distribution \citet{Uhlemann2023}  showed that the large-field approximation (in their notation $P_d(\theta)\simeq \theta$) leads to model covariances in agreement with simulations, and so the SSC is less important for their statistic. This indicates that the impact of SSC depends on the considered statistic.

%
%

\begin{acknowledgements}
  We thank the anonymous referee, as well as Alex Barreira and Alex Hall, for their thoughtful and valuable comments. Funded by the TRA Matter (University of Bonn) as part of the Excellence Strategy of the federal and state governments. This work has been supported by the Deutsche Forschungsgemeinschaft through the project SCHN 342/15-1 and DFG SCHN 342/13. PAB and SH acknowledge support from the German Academic Scholarship Foundation. LP acknowledges support from the DLR grant 50QE2002. We would like to thank Joachim Harnois-D\'eraps for making public the SLICS mock data, which can be found at \url{http://slics.roe.ac.uk/}. We thank Oliver Friedrich and Niek Wielders for helpful discussions.
\end{acknowledgements}

%
%

\bibliography{cite}

%

\begin{appendix}
  \onecolumn 
\section{\label{app: covariance calculation}\texorpdfstring{Expressing $C_{\hat{\Xi}}$ in terms of the power- and trispectrum}{Expressing covariance in terms of the power spectrum and trispectrum}}

In this Appendix, we relate $C_{\hat{\Xi}}$ to the power spectrum $P$ and the trispectrum $T$. We start from Eq.~\eqref{eq: covariance zeroth step} and rewrite it in terms of the Fourier transform $\Tilde{\kappa}$ of the convergence, using
\begin{equation}
    \int \dd[2]{\vartheta} \kappa(\varthetavec) \, U_{1/2}(\theta, \varthetavec-\alphavec) = \int \frac{\dd[2]{\ell}}{(2\pi)^2}\Tilde{\kappa}(\ellvec)\, \Tilde{U}_{1/2}(\theta, \ellvec)\;.
\end{equation}
This leads to
\begin{align}
    \expval{\hat{\Xi}(\theta_1)\, \hat{\Xi}(\theta_2)} &= \frac{1}{A^2}\int \dd[2]{\alpha_1} \int \dd[2]{\alpha_2} W_A(\alphavec_1)\, W_A(\alphavec_2)\,\int \frac{\dd[2]{\ell}_1}{(2\pi)^2} \int \frac{\dd[2]{\ell}_2}{(2\pi)^2} \int \frac{\dd[2]{\ell}_3}{(2\pi)^2} \int \frac{\dd[2]{\ell}_4}{(2\pi)^2}\\
    &\notag \quad \times \tilde{U}_1(\theta_1, \ellvec_1)\,\tilde{U}_2(\theta_1, \ellvec_2)\, \tilde{U}_1(\theta_2, \ellvec_3)\, \tilde{U}_2(\theta_2, \ellvec_4)\, \, \E^{\I (\ellvec_1+\ellvec_2)\alphavec_1+\I (\ellvec_3+\ellvec_4)\alphavec_2} \,  \expval{\Tilde{\kappa}(\ellvec_1)\,\Tilde{\kappa}(\ellvec_2)\,\Tilde{\kappa}(\ellvec_3)\,\Tilde{\kappa}(\ellvec_4)}
\end{align}
We decompose the four-point function as suggested by Eq.~\eqref{eq: split kappa four point}. This leads to
\begin{align}
    \expval{\hat{\Xi}(\theta_1)\,\hat{\Xi}(\theta_2)} &= \frac{1}{A^2}\int \dd[2]{\alpha_1} \int \dd[2]{\alpha_2} W_A(\alphavec_1)\, W_A(\alphavec_2)\,\int \frac{\dd[2]{\ell}_1}{(2\pi)^2} \int \frac{\dd[2]{\ell}_2}{(2\pi)^2} \int \frac{\dd[2]{\ell}_3}{(2\pi)^2} \int \frac{\dd[2]{\ell}_4}{(2\pi)^2}\\
    &\notag \quad \times \tilde{U}_1(\theta_1, \ellvec_1)\,\tilde{U}_2(\theta_1, \ellvec_2)\, \tilde{U}_1(\theta_2, \ellvec_3)\, \tilde{U}_2(\theta_2, \ellvec_4)\, \, \E^{\I (\ellvec_1+\ellvec_2)\alphavec_1+\I (\ellvec_3+\ellvec_4)\alphavec_2} \\
    &\notag \quad \times \Bigg[ P(\ell_1)\,P(\ell_3)\, (2\pi)^4\, \dirac(\ellvec_1+\ellvec_2)\, \dirac(\ellvec_3+\ellvec_4) + P(\ell_1)\,P(\ell_2)\, (2\pi)^4\, \dirac(\ellvec_1+\ellvec_3)\, \dirac(\ellvec_2+\ellvec_4) \\
    &\notag \qquad + P(\ell_1)\,P(\ell_2)\, (2\pi)^4\, \dirac(\ellvec_1+\ellvec_4)\, \dirac(\ellvec_2+\ellvec_3) + T(\ellvec_1, \ellvec_2, \ellvec_3, \ellvec_4) (2\pi)^2\, \dirac(\ellvec_1+\ellvec_2+\ellvec_3+\ellvec_4)\Bigg]    
\end{align}
Given that
\begin{equation}
    \expval{\hat{\Xi}(\theta)}=\frac{1}{A} \int \dd[2]{\alpha} W_A(\alpha) \int \frac{\dd[2]{\ell}}{(2\pi)^2} P(\ell)\, \tilde{U}_1(\theta, \ellvec)\,\tilde{U}_2(\theta, \ellvec)
\end{equation}
the covariance can be written
\begin{align}
    C_{\hat{\Xi}}(\theta_1, \theta_2) &= \frac{1}{A^2}\int \dd[2]{\alpha_1} \int \dd[2]{\alpha_2} W_A(\alphavec_1)\, W_A(\alphavec_2)\, \Bigg[ \int \frac{\dd[2]{\ell}_1}{(2\pi)^2} \int \frac{\dd[2]{\ell}_2}{(2\pi)^2} \E^{\I (\ellvec_1+\ellvec_2)(\alphavec_1-\alphavec_2)}   P(\ell_1)\,P(\ell_2)\\
    &\notag \quad \times \left(\tilde{U}_1(\theta_1, \ellvec_1)\,\tilde{U}_2(\theta_1, \ellvec_2)\, \tilde{U}_1(\theta_2, \ellvec_1)\, \tilde{U}_2(\theta_2, \ellvec_2)+\tilde{U}_1(\theta_1, \ellvec_1)\,\tilde{U}_2(\theta_1, \ellvec_2)\, \tilde{U}_1(\theta_2, \ellvec_2)\, \tilde{U}_2(\theta_2, \ellvec_1)\right) \\
    &\notag \quad +\int \frac{\dd[2]{\ell}_1}{(2\pi)^2} \int \frac{\dd[2]{\ell}_2}{(2\pi)^2} \int \frac{\dd[2]{\ell}_3}{(2\pi)^2}  \E^{\I (\ellvec_1+\ellvec_2)(\alphavec_1-\alphavec_2)}\, T(\ellvec_1, \ellvec_2, \ellvec_3, -\ellvec_1-\ellvec_2-\ellvec_3) \\
    &\notag \quad \times \tilde{U}_1(\theta_1, \ellvec_1)\,\tilde{U}_2(\theta_1, \ellvec_2)\, \tilde{U}_1(\theta_2, \ellvec_3)\, \tilde{U}_2(\theta_2, \ellvec_1+\ellvec_2+\ellvec_3)\Bigg]  \\
    &=\int \frac{\dd[2]{\ell}_1}{(2\pi)^2} \int \frac{\dd[2]{\ell}_2}{(2\pi)^2} G_A(\ellvec_1+\ellvec_2)\,  P(\ell_1)\,P(\ell_2)\\
    &\notag \quad \times \left(\tilde{U}_1(\theta_1, \ellvec_1)\,\tilde{U}_2(\theta_1, \ellvec_2)\, \tilde{U}_1(\theta_2, \ellvec_1)\, \tilde{U}_2(\theta_2, \ellvec_2)+\tilde{U}_1(\theta_1, \ellvec_1)\,\tilde{U}_2(\theta_1, \ellvec_2)\, \tilde{U}_1(\theta_2, \ellvec_2)\, \tilde{U}_2(\theta_2, \ellvec_1)\right) \\
    &\notag \quad +\int \frac{\dd[2]{\ell}_1}{(2\pi)^2} \int \frac{\dd[2]{\ell}_2}{(2\pi)^2} \int \frac{\dd[2]{\ell}_3}{(2\pi)^2}  G_A(\ellvec_1+\ellvec_2)\\
    &\notag \quad \times T(\ellvec_1, \ellvec_2, \ellvec_3, -\ellvec_1-\ellvec_2-\ellvec_3)\,\tilde{U}_1(\theta_1, \ellvec_1)\,\tilde{U}_2(\theta_1, \ellvec_2)\, \tilde{U}_1(\theta_2, \ellvec_3)\, \tilde{U}_2(\theta_2, \ellvec_1+\ellvec_2+\ellvec_3)\;,
\end{align}
where we introduce the geometry factor $G_A$ defined in Eq.~\eqref{eq: definition GA}.

\section{\label{app: validation}Details on validation measurements}
 All calculations in this work are performed for a flat $\Lambda$CDM cosmology with dimensionless Hubble-parameter $h=0.69$, clustering parameter $\sigma_8=0.83$, matter density $\Omm=0.29$, baryon density $\Omega_\mathrm{b}=0.047$, and power spectrum scale index $n_\mathrm{s}=0.969$. 
 
\subsection{\label{app: validation: data}Validation data}
We use the same mock shear catalogues from the Scinet LIghtcone Simulations (SLICS, \citealp{HarnoisDeraps2015}) as used in \citet{Heydenreich2023} and \citet{Linke2023a}. The simulations use a flat $\Lambda$CDM cosmology with our fiducial cosmological parameters and contain $1536^3$ particles inside a $505\, h^{-1}\,$Mpc box. We use shear catalogues from $924$  pseudo-independent lines of sight, each with a square area of $100\,\mathrm{deg}^2$. The source galaxies are distributed with a redshift distribution of
\begin{equation}
    n(z) \propto z^2\, \E^{-\left(z/z_0\right)^\beta}\;,
\end{equation}
with $z_0=0.637$, $\beta=1.5$, and normalisation such that the overall galaxy density is $30\, \mathrm{arcmin}^{-2}$, as expected for a Stage IV lensing survey. The shear includes shape noise, which is infused by adding random ellipticities from a Gaussian distribution. The two-component ellipticity dispersion $\sigma_\epsilon^2$ of the shape noise is $(0.37)^2$.

\subsection{\label{app: validation: measurement}Covariance measurement}
We estimate the covariance in the simulation in two different ways. In the first approach, we replace the expectation value of $\MapMap$ by its spatial average for each realisation of the simulation and use the sample covariance between the realisation as covariance estimate. To do so, we use that the aperture mass filter $U_\theta$ is related to a filter $Q_\theta$, for which
\begin{equation}
    \label{eq: Map3 from gamma_t}
    \Map( \theta; \varthetavec)=\int \dd[2]{\vartheta'}\; Q_\theta(|\varthetavec - \varthetavec'|)\, \gamma_\mathrm{t}(\varthetavec';\varthetavec)\;,
\end{equation}
where $Q_\theta$ is given by
\begin{equation}
    Q_\theta(\vartheta)=\frac{2}{\vartheta^2}\, \int_0^\vartheta \dd{\vartheta'}\; \vartheta'\, U_\theta(\vartheta') - U_\theta(\vartheta)\;.
\end{equation}
We calculate Eq.~\eqref{eq: Map3 from gamma_t} using Fast Fourier Transform (FFT) for the aperture scale radii $\theta \in \{\astroang{;4;}, \astroang{;8;}, \astroang{;16;}\}$, which gives us an aperture mass map ${\Map}^{(i)}(\varthetavec, \theta)$ for each realisation $i$ and scale radius $\theta$. To remove border effects, we cut off a border of four times the largest aperture radius (i.e. $4\times\astroang{;16;}=\astroang{;64;}$) from each side of the aperture mass maps. Then we square the aperture mass maps for each scale radius and subsequently take the average over all pixel values to get an estimate for $\MapMap$ for each realisation. We take the sample covariance of these realisations as our covariance estimate.

For the sample covariance, we estimate uncertainties using bootstrapping. For this, we create $10\,000$ lists of 924 randomly drawn integers between 1 and 924, where each list can contain any integer multiple times. We calculate a sample covariance for each list using the $\MapMap^{(i)}$ for which $i$ is a list element, leading to $10\,000$ covariance estimates. Our uncertainty estimate is the standard deviation of these $10\,000$ estimates.

The second approach to estimate $C_{\MapMap}$ in the simulations uses Eq.~\eqref{eq: covariance from correlation functions}. For this, we use the squared aperture mass maps created before. We then calculate the correlation function $\xi^{(i)}_{\Map^2}$ for each realisation $i$ using 
\begin{equation}
\xi^{(i)}_{\Map^2}(\theta_1, \theta_2, \etavec) =     \frac{1}{A}\frac{1}{E_A(\etavec)}\,\int_A \dd[2]{\alphavec} {\Map^{(i)}}^2(\theta_1, \alphavec)\,{\Map^{(i)}}^2(\theta_2),\alphavec+\etavec\, W(\alphavec+\etavec)\;
\end{equation}
We evaluate this equation using FFT, as implemented in the \verb|scipy| routine \verb|correlate|. The mean of all $\xi^{(i)}$ is then plugged into Eq.~\eqref{eq: covariance from correlation functions} to yield the covariance estimate $C_{\MapMapEst}^\mathrm{corr}$.

We measure $\xi_{\MapMap}$ as a function of the two-dimensional vector $\etavec$, not just its magnitude $\eta$. Therefore, the measurement method allows us to test the statistical anisotropy of the convergence field by comparing, for example, $\xi_{\MapMap}(\etavec)$ with $\xi_{\MapMap}(-\etavec)$.

\subsection{\label{app: validation: modelling}Polyspectra modelling}
We model the covariance once with the large-field approximation $C_{\hat{\Xi}}^\infty$ from Eq.~\eqref{eq: covariance large-field approximation} and once including the SSC term in Eq.~\eqref{eq: covariance with ssc}. For this, we use the revised \verb|halofit| prescription for the power spectrum \citep{Takahashi2012}. The trispectrum is modelled with the one-halo term of the halo model \citep{Cooray2002}, using a Sheth--Tormen halo mass function and halo bias \citep{Sheth1999}, and Navarro--Frenk--White halo profiles \citep{Navarro1996}.

\end{appendix}

\end{document}